\documentclass[a4paper,
               keeplastbox,   
               ]{jacow}
\usepackage{pdfpages,multirow,ragged2e} %
\makeatletter%
	\ifboolexpr{bool{xetex}}
	 {\renewcommand{\Gin@extensions}{.pdf,%
	                    .png,.jpg,.bmp,.pict,.tif,.psd,.mac,.sga,.tga,.gif,%
	                    .eps,.ps,%
	                    }}{}
\makeatother

\ifboolexpr{bool{xetex} or bool{luatex}} 
 {}                                      
 {\usepackage[utf8]{inputenc}}           

\usepackage[USenglish]{babel}

\ifboolexpr{bool{jacowbiblatex}}%
 {%
  \addbibresource{jacow-test.bib}
  \addbibresource{biblatex-examples.bib}
 }{}
\begin{document}

\title{Implementation of synchronised PS-SPS transfer with barrier bucket}

\author{M. Vadai\thanks{mihaly.vadai@cern.ch}, H. Damerau, M. Giovannozzi, A. Huschauer, A. Lasheen, CERN, Geneva, Switzerland}
	
\maketitle

\begin{abstract}
For the future intensity increase of the fixed-target beams in the CERN accelerator complex, a barrier-bucket scheme has been developed to reduce the beam loss during the 5-turn extraction from the PS towards the SPS, the so-called Multi-Turn Extraction. The low-level RF system must synchronise the barrier phase with the PS extraction and SPS injection kickers to minimise the number of particles lost during the rise times of their fields. As the RF voltage of the wide-band cavity generating the barrier bucket would be too low for a conventional synchronisation, a combination of a feedforward cogging manipulation and the real-time control of the barrier phase has been developed and tested. A deterministic frequency bump has been added to compensate for the imperfect circumference ratio between PS and SPS. This contribution presents the concept and implementation of the synchronised barrier-bucket transfer. Measurements with high-intensity beam demonstrate the feasibility of the proposed transfer scheme.
\end{abstract}

\section{introduction}

The Multi-Turn Extraction (MTE) scheme~\cite{MTE2001} replaced the previous Continuous Transfer (CT) extraction method~\cite{CT1973, CT2011, MTEEPL} in the CERN Proton Synchrotron (PS) to deliver the high-intensity proton beams for the fixed-target physics at the CERN Super Proton Synchrotron (SPS), and the operational implementation of MTE allowed to significantly reduce extraction losses in the PS ring  (see, e.g. ~\cite{MTEfluct,MTEPRAB,MTEFeatures}). 

The circumference difference between the PS and the SPS, the latter eleven times longer than the first, suggests extracting the beam from the PS over five turns to maximise the duty factor for fixed-target experiments. This means two transfers from the PS which fills 10/11th of the SPS and leaves time for a gap for the injection kickers. MTE has been designed to split the beam into five beamlets in the horizontal plane by crossing adiabatically the fourth-order resonance~\cite{MTE2001,MTEPRE}. Only losses due to the longitudinal beam structure and the rise time of the extraction kickers remain.

The intense studies on the theory behind transverse beam splitting~\cite{MTEPRE} had ensured that the PS performance is close to an optimum. Studies~\cite{MTEHB18,MTEHI} confirmed that no showstopper is in sight when increasing intensity for fixed-target physics at the SPS. 


The remaining extraction beam losses are a consequence of the SPS requirements for the beam transfer from the PS, as a debunching is performed after transverse beam splitting and prior to extraction. While this is necessary for the SPS, it is certainly a drawback for the PS, given that the non-negligible rise time of the extraction kickers induces beam losses during beam extraction. A much more promising approach appears to be the implementation of a barrier bucket~\cite{Foldy,Griffin,BB_particle_dynamics,AGSbarrierbucket,BB_multiple_inj,Takayama-induction} which has the potential to practically remove the extraction losses in PS. The advantage of this approach is that no new hardware is required, as a wideband RF cavity is already present in the PS ring. It is loaded with Finemet$^\circledR$ material, which makes it usable in the frequency range from $400$~kHz to well above $10$~MHz. This device was installed in 2014~\cite{Finemet,PSFinemet} as a part of the longitudinal coupled-bunch feedback system. 

Initial beam tests pursued at the PS gave extremely encouraging results~\cite{BBlineartech,BBMTEEPL,BBMTEIOP}. It has been possible to successfully implement barrier-bucket manipulations in the PS, and even combine it with the transverse beam splitting yielding a substantial reduction of beam losses at PS extraction (see, e.g.~\cite{prab-barrier-bucket} for a review of this beam manipulation). 

\section{synchronisation concept}

\begin{figure}[!tbh]
    \centering
    \includegraphics*[width=\columnwidth]{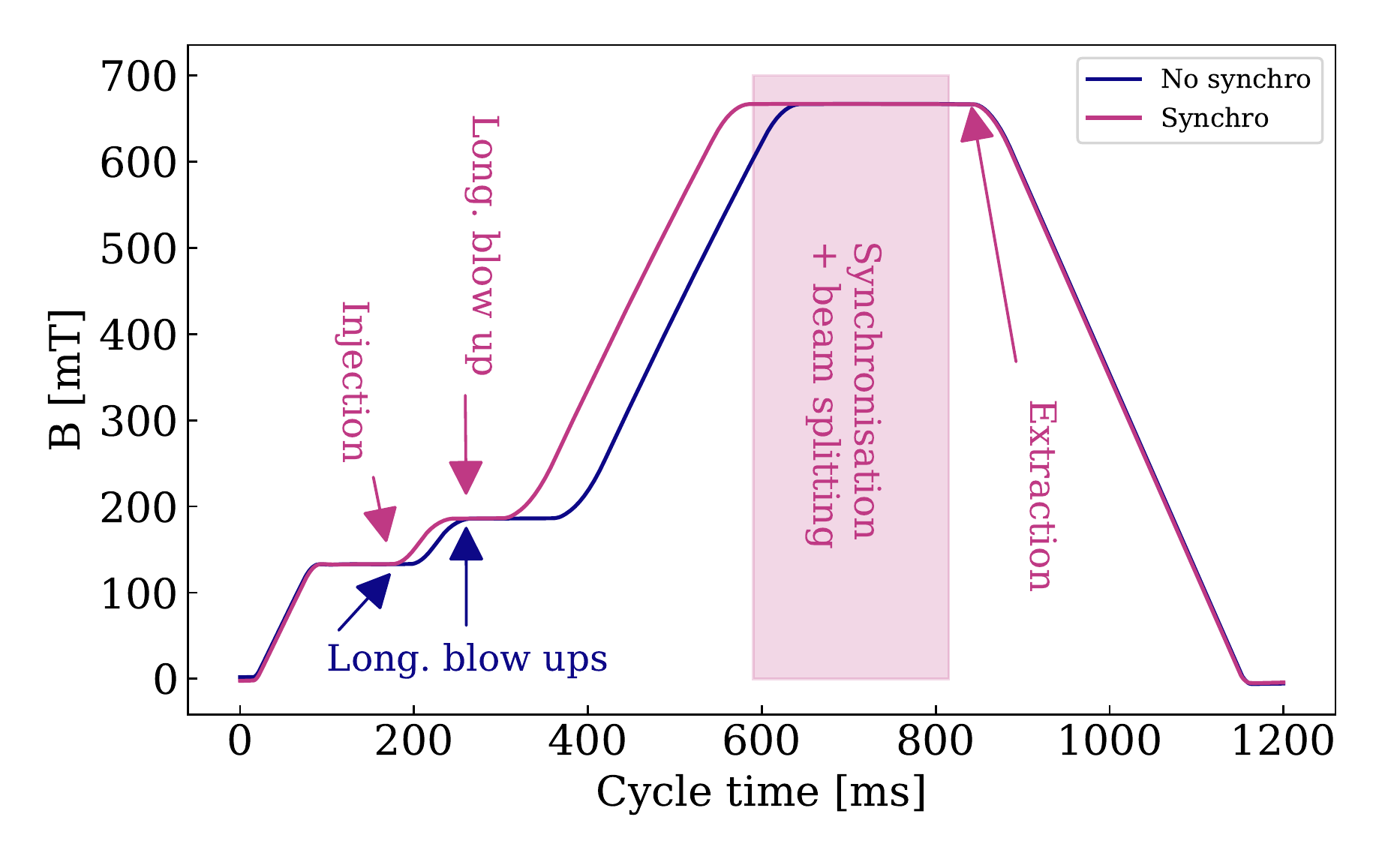}
    \caption{A comparison of the magnetic field function of the operational SFTPRO cycle without the synchronization and the modified cycle with the longer flat-top needed for synchronisation.}
    \label{fig:B-train}
\end{figure}

Figure~\ref{fig:B-train} shows the magnetic flux density during the acceleration cycle with and without the synchronisation process. After injection, the beam is accelerated in $h=8$ buckets, then longitudinally blown up, split into $h=16$ buckets and accelerated to the flat-top momentum of 14 GeV/$c$. The longitudinal blow-up is needed to stabilise the beam at transition crossing at high intensities. The bunches are then transversely split and debunched at the end of the cycle prior to extraction to the SPS.

The transfer of the coasting beam from the PS to the SPS is triggered by the SPS, which means that the extraction kickers of the PS are synchronous with the injection kickers of the SPS. In order to mitigate the losses caused by the rise time of the PS extraction kickers, a longitudinal gap is made with the wide-band RF system of the PS. This new requirement means that RF systems in PS and SPS that were previously operated stand-alone must be synchronised. The beam energy has to be temporarily changed in the PS to overlap the position of the longitudinal gap with the rise time of the PS extraction kickers. This can only be done with enough voltage available in the main RF system, especially if the manipulation is to be performed on the order of ten milliseconds. 

Due to the low RF voltage requirements of the transverse beam splitting, $V_\mathrm{RF}$ must be lowered well before the end of the cycle. The Finemet system can only generate a low RF voltage, too small to synchronise the beam. Hence the synchronisation is proposed to take place at the beginning of the magnetic flat-top, where the RF voltage is still sufficient to re-phase the beam quickly. Due to the periodicity it only requires at maximum a half of a $h=16$ bucket phase change in either direction. This is also required because the barrier-bucket transfer from $h=16$ is non-adiabatic, which means that the barrier is most effective once the voltage is already raised between two existing $h=16$ buckets, as shown in Fig.~\ref{fig:longitudinal-profiles}.

The beam phase measurements for $h=16$ and then $h=1$ have to take place at the same frequency as the SPS. At a fixed bending field in the PS this means that the mean radial position at the time of the measurements must be the same as the one at extraction. This poses a problem for the transverse splitting process as its ideal radial position is centred. Hence, in open loop, a constant frequency excursion is programmed to centre the beam for the transverse splitting and then after the process steer it to the extraction orbit with the RF systems. This only introduces a fixed, repeatable phase offset that does not affect the synchronisation.

Since the bunches after the cogging are $h=16$ synchronous with the SPS, the final $h=1$ synchronisation is only a bucket selection, which can be pre-calculated. Therefore, no change in beam energy is needed for the $h=1$ part.

The steps of the proposed synchronisation sequence are summarised in Table~\ref{l2ea4-t1}.

\begin{table}[!hbt]
   \centering
   \caption{Synchronisation sequence in PS}
   \begin{tabular}{ll}
       \toprule
     
       \textbf{Cycle time [\SI{}{ms}]} & \textbf{Action}                      \\
       \midrule
            590-600       &  Loops off, $h=16$ phase measurement    \\
            600     & $h=1$ phase measurement          \\ 
            600-620     & Cogging           \\ 
            620-650        & Radial steering to central orbit        \\
            650-770        & Constant offset for MTE              \\
            770-800        & Radial steering to extraction orbit            \\
            800-815     & Hand-over $h=16$ to barrier bucket         \\
            800-815     & Debunching              \\
            835     & Extraction              \\
       \bottomrule
   \end{tabular}
   \label{l2ea4-t1}
\end{table}

\subsection{Cogging}

The principle of phase correction at a fixed magnetic field, $B$ is to change the frequency of the beam such that the integral of the frequency change equals the desired phase difference. This requires a phase measurement at the reference frequency, a phase correction, and then returning to the original frequency, thus implying that the beam has to be accelerated and then decelerated.

Since the bending field is constant, the beam is also radially offset according to the relation~\cite{Baudrenghien:EPAC98-WEP02H,Bovet-formulae}:

\begin{equation}
    \frac{df}{f} = \frac{\gamma^2}{\gamma_{\rm tr}^2-\gamma^2} \frac{dR}{R} \label{eq:dfoverf} \ ,
\end{equation}
where $f$ is the beam revolution frequency, $R$ the PS radius, $\gamma$ the Lorentz factor, and $\gamma_\mathrm{tr}$ the gamma at transition. The frequency offset needs to be small enough due to aperture limitation. Longitudinal macro-particle tracking simulations have been performed to validate the phase curve~\cite{phd}
\begin{eqnarray}
\phi(\phi_{\mathrm{set}}, T, t) & = &  \frac{\phi_{\mathrm{set}}}{T}\left[t - \frac{T}{2\pi}\sin\left(\frac{2\pi}{T}t\right) \right] \\
t \in [0,T] \ & ; & \ \phi_{\mathrm{set}} \in [-2 \pi, 2 \pi] \ ,
\end{eqnarray}
which corresponds to the programmed frequency curve of
\begin{equation}
f (\phi_{\mathrm{set}}, T, t) = \frac{\phi_{\mathrm{set}}}{2\pi T}\left[1 - \cos\left(\frac{2\pi}{T}t\right) \right] \label{eq:frequency} \ .
\end{equation}
It defines the set of frequency curves for the $h=16$ correction, which can be seen in Fig.~\ref{fig:frev-mrp}~(middle). Note that the phase range is set conservatively for a whole bucket movement in either direction, which is useful for testing - half of this phase change is sufficient in operation. 

\section{Implementation}

\begin{figure}[!tbh]
    \centering
    \includegraphics*[trim = 0mm 0mm 0mm 0mm, width=\columnwidth]{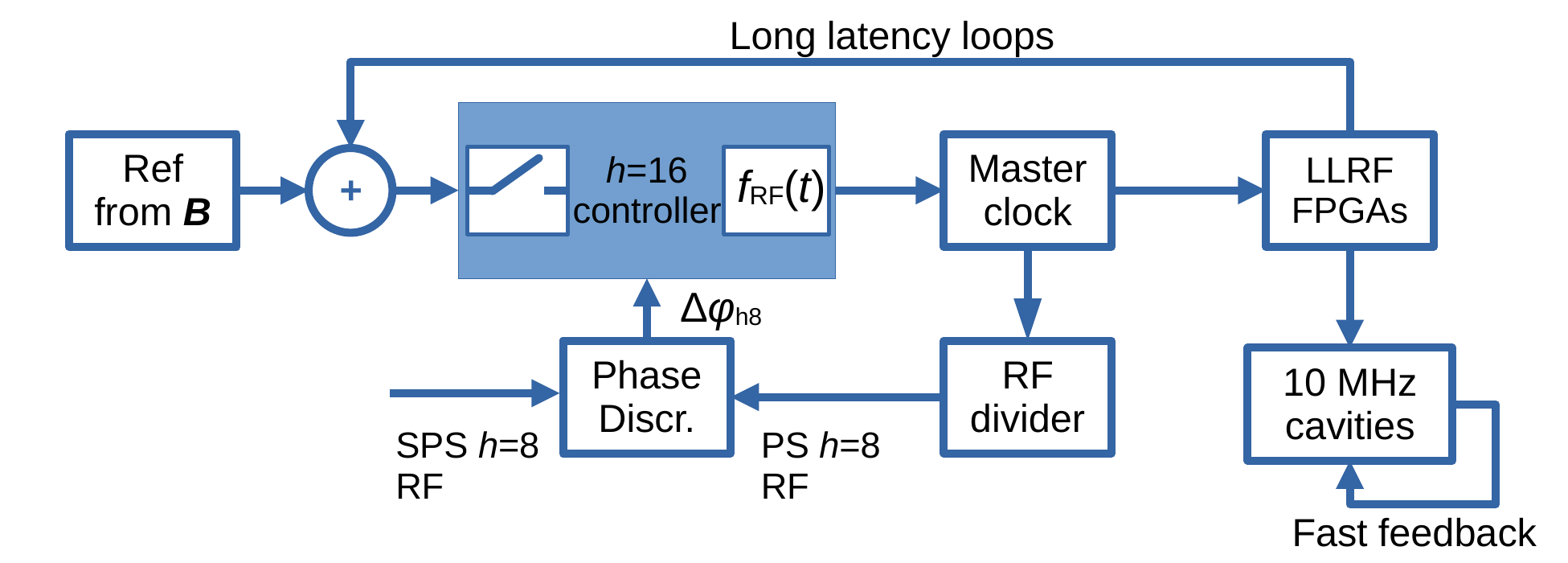}
    \caption{$h=16$ cogging configuration.}
    \label{fig:h16scheme}
\end{figure}

Synchronisation can be tried with the current PS beam control system if one acts on the master direct digital synthesizer~(DDS) frequency as depicted in Fig.~\ref{fig:h16scheme}. Phase slips and frequency steering can be implemented for all RF systems involved in the synchronisation simultaneously since the master clock frequency drives the clock signal for the low-level RF~(LLRF) cavity controllers. Since this is an open-loop manipulation, the handover from closed-loop~(see long latency loops in Fig.~\ref{fig:h16scheme}) to open-loop has to be implemented. In order to significantly reduce phase drift during flat-top, the precision of the frequency word has to be increased from~23 bits to 32 bits in open loop.

\begin{figure}[!tbh]
    \centering
    \includegraphics*[width=\columnwidth]{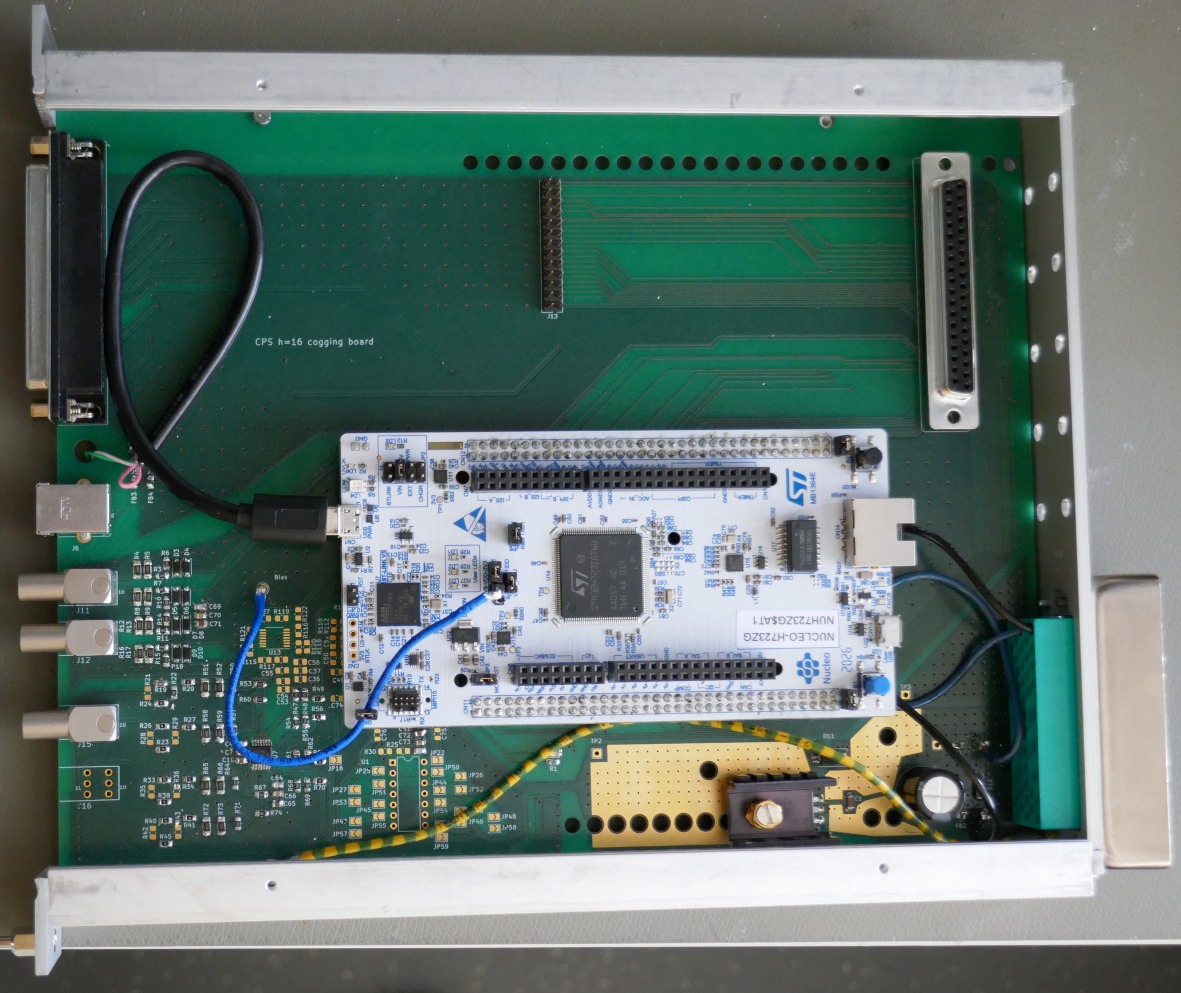}
    \caption{AED-00299 prototype module to evaluate the $h=16$ synchronisation feature with beam using the current control system in the PS (designed with the open source KiCAD~\cite{kicad}).}
    \label{fig:h16hardware}
\end{figure}

To reduce development time and cost, a rapid prototyping solution was chosen, also taking into account the challenge of electronic component availability. A custom board was designed to interface an ARM\circledR Cortex\circledR M7 controller~(STM32H723ZG) on a development board with the current beam control (see Fig.~\ref{fig:h16hardware}). The controller's digital interface is compatible with the LVTTL/TTL signals of the frequency distribution~\cite{Sladen:734627}. Hence, adding only a thin layer of power management and IO interface was sufficient. The analogue part of the $h=16$ board contains operational amplifier~(ADA4891) based circuits suitable for interfacing RF signals from DC to \SI{22}{MHz} or DC to $h=46$ with the 16 bit ADCs of the controller.

\begin{figure}[!tbh]
    \centering
    \includegraphics*[width=\columnwidth]{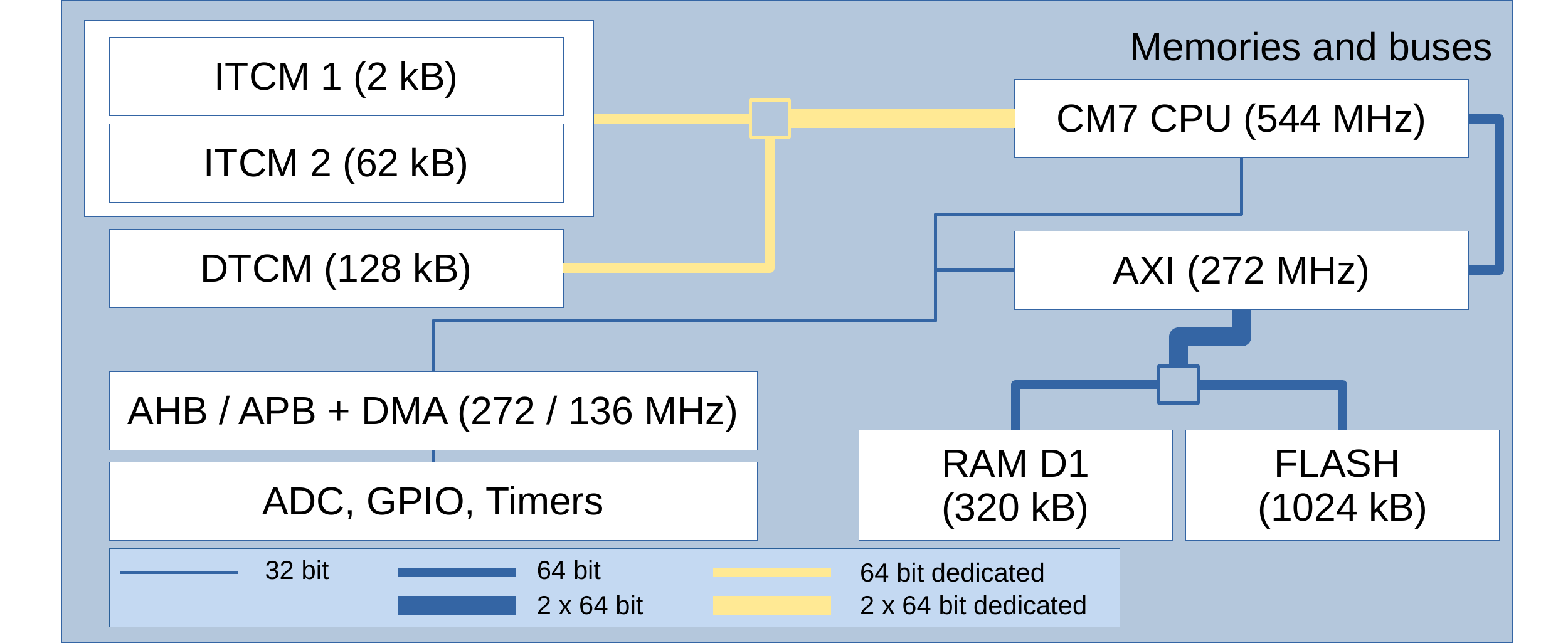}
    \caption{The parts of the STM H7 memory and bus architecture used in the implementation.}
    \label{fig:firmware-structure}
\end{figure}

At the micro-controller architecture level~(see Fig.~\ref{fig:firmware-structure}), the CPU-intensive low-latency processing uses the Instruction and Data Tightly Coupled Memories~(ITCM/DTCM). The peripheral connections to the advanced high performance bus~(AHB) that do not require immediate CPU action are handled by the Direct Memory Access~(DMA) controller. The corresponding data is stored in the slower RAM without interrupting the CPU. The Advanced Peripheral Bus~(APB) connects the architecture's advanced timers used for the phase measurement. Therefore, the CPU is not required for the measurement itself. 
 
\begin{figure}[!tbh]
    \centering
    \includegraphics*[width=\columnwidth]{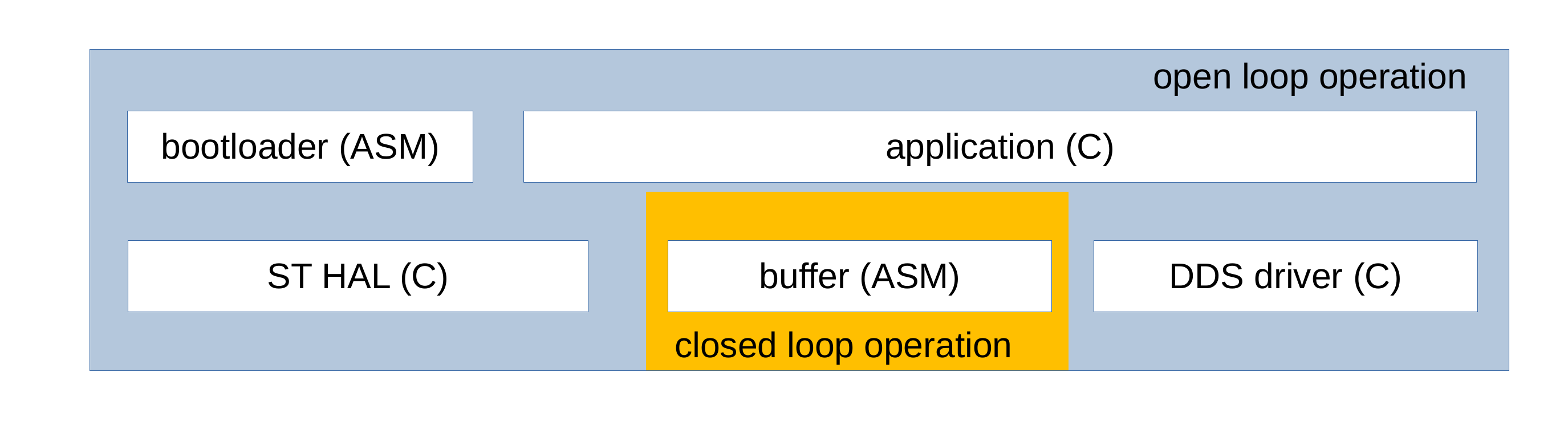}
    \caption{The structure of the firmware with the programming languages used (ASM for ARM Assembler and C).}
    \label{fig:software-structure}
\end{figure}

The software architecture (see Fig.~\ref{fig:software-structure}) was kept very simple in order to implement only the strictly necessary components. No operating system is needed for the synchronisation-related functions, since a reliable and predictable scheduling can be achieved by ARM's Nested Vector Interrupt Controller hardware~(NVIC).

\subsection{Frequency word generation}

The implementation of Eq.~\eqref{eq:frequency} in discrete time taking the architecture of the controller into account takes the following form:

\begin{eqnarray}
T = N\tau_\mathrm{s} \ & ; & \ t = n\tau_\mathrm{s} \\
f (\phi_{\mathrm{set}}, N, \tau_\mathrm{s}, n) & = & \frac{h_{\mathrm{CLK}}}{h_{\mathrm{RF}}} \frac{ \phi_{\mathrm{set}}}{2\pi N\tau_\mathrm{s}}\left[1 - \cos\left(\frac{2\pi}{N}n\right) \right] \label{eq:freq-impl} ,
\end{eqnarray}
where $h_{\mathrm{CLK}} = 256$, $h_{\mathrm{RF}} = 16$, $N$ is the number of points stored in memory, $n$ is the running sample index, $\tau_\mathrm{s}$ is the sampling period. It can be seen that the cosine term on the right-hand side of Eq.~\eqref{eq:freq-impl} does not depend on time or the chosen sampling frequency, and it therefore does not need to be recalculated even if the frequency-bump parameters, such as the length or the desired phase offset change. The parameter that typically changes from cycle to cycle is $\phi_{\mathrm{set}}$, which is the result of the phase measurement. This means that only one multiplication per output sample must be performed in real time in addition to the encoding, and that the multiplication takes only one clock cycle at \SI{544}{MHz} clock frequency. 

Furthermore, to test the limit of adiabaticity of beam manipulations, the sampling time $\tau_\mathrm{s}$ can be automatically adjusted if a very fast ($< \SI{5}{ms}$) frequency bump has to be performed. This change, again, does not affect the cosine term in Eq.~\eqref{eq:freq-impl}, since the samples are simply skipped and only the scaling factor is adjusted. The trade-off is that the precision at which the curve approximates the desired phase change is lower when several samples are skipped.

\subsection{Phase measurements and barrier-bucket selection}


The $h=16$ phase measurement is performed reading the output of an existing $h=16$ phase discriminator in the PS as shown in Fig.~\ref{fig:h16scheme}. This phase discriminator has a dead zone, but it is sufficiently small that it only causes a very rare poor measurement, approximately one every hour or so. However, if one were to implement a $h=1$ measurement in the same way, then the available equipment would have a much larger dead zone.

\begin{figure}[!tbh]
    \centering
    \includegraphics*[trim = 10mm 10mm 10mm 10mm, width=\columnwidth]{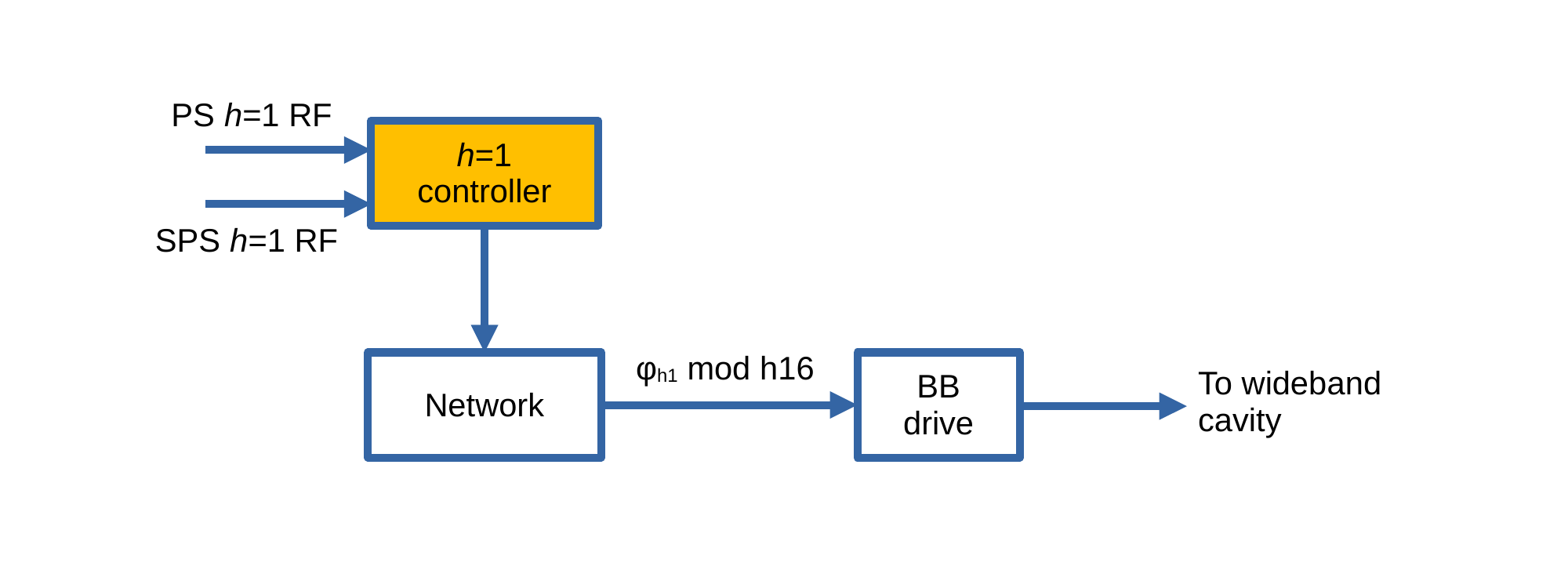}
    \caption{$h=1$ phase measurement is performed by the controller directly. The result is sent over the network in real time to generate the potential barrier at the correct azimuth within the same cycle.}
    \label{fig:h1scheme}
\end{figure}

Hence, a dedicated measurement for the $h=1$ bucket selection (see Fig.~\ref{fig:h1scheme}) was developed. This requires only a minimal hardware interface in the form of a development board hat~(AED-00300). Using a combination of advanced timers on the controller, the gap selection can be achieved with a precision at the degree level. This implementation does not have a dead zone, which would result in the barrier being at a completely different azimuth location compared to the target. The accuracy of the scheme is evaluated with beam in the next section.

\section{Validation with beam}

\begin{figure}[!tbh]
    \centering
    \includegraphics*[width=\columnwidth]{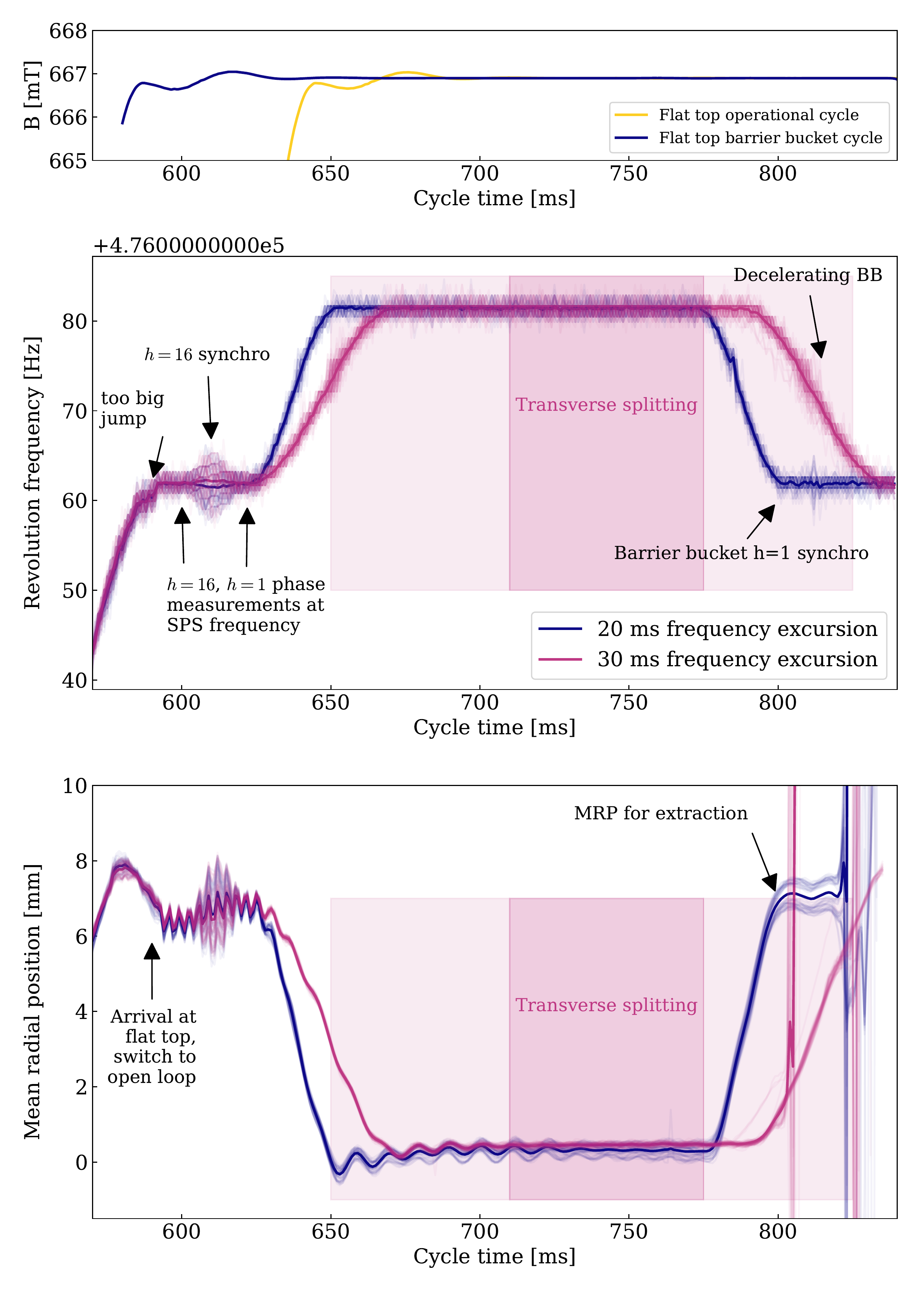}
    \caption{The $B$-field (top), the revolution frequency (middle), and mean radial position (MRP) (bottom) evolution during a machine development session to set up the synchronisation process in July 2022. The plots highlight potential pitfalls and their effects. Mitigation measures are explained in the text. Data from 132 cycles are shown for two different values of the time for the frequency excursions.}
    \label{fig:frev-mrp}
\end{figure}

As described above, the synchronisation occurs at fixed bending field, as indicated by Fig.~\ref{fig:frev-mrp}~(top). The plot was generated using measured data with beam during setup, hence it also shows potential pitfalls with their mitigation measures as described below.

The frequency at the flat-top is fixed at first, and the loops are opened. The radial position has to be carefully matched such that there is no sudden jump of the programmed fixed frequency corresponding to the SPS reference frequency, and the closed loop frequency of the PS. If there is a considerable difference, phase oscillations visible on Fig.~\ref{fig:frev-mrp}~(bottom) before \SI{610}{ms} are triggered. If the radial positions are matched to about \SI{0.1}{mm} using radial loop corrections before switching to the open loop, the smooth switch in the firmware can bridge the small remaining difference. However, these oscillations must be corrected to conserve the longitudinal emittance and to keep the radial position at extraction within acceptable limits, especially with increasing intensity.

The phase is measured with respect to the SPS reference, in units of $h=16$ at fixed frequency, and a correction with the main RF system is performed, see Fig.~\ref{fig:frev-mrp}~(middle). This correction varies from cycle to cycle as this is the main action to synchronise the beams. Therefore, the overlapping graphs~(in the middle and bottom plots) from different cycles show the envelope of all the corrections between~600~ms~and~620~ms in the cycle. Then the $h=1$ phase is measured with respect to the SPS and this information is sent to the barrier-bucket front end over the network during transverse beam splitting. The beam is brought to a favourable radial position for transverse splitting, with a constant phase difference between PS and SPS. Oscillations of the mean radial position (MRP) at the beginning of the transverse beam splitting decay during the process. Since the splitting itself occurs towards the end of the process, the effect of these is insignificant. Once transversely split, the beam is brought to back to the SPS reference frequency. At this point, the barrier front end received the correct barrier phase over the network and selected the correct gap between the $h=16$ buckets to raise the barrier.

Two barrier-bucket voltage ramping schemes were tried. The traces marked in purple in Fig.~\ref{fig:frev-mrp}~(middle) and~(bottom) show a case when the barrier voltage is increased while the final frequency excursion in the PS is still performed. This effectively results in a decelerating barrier-bucket configuration. The advantage of the decelerating barrier is that the frequency change can be spread out in time, but has to be tuned to the longitudinal dynamics during debunching to conserve the depth of the gap. This scheme is currently under study, it is not stable for trial runs yet. The barrier at fixed position was used for further tests because of its simplicity.


The detailed quantitative consequences of the longitudinal emittance on the losses at extraction and the limits of the scheme in terms of beam intensity are subject to studies. Current tests were carried out up to about 15\% less than nominal intensity of~$\SI{2.1}{\times 10^{13} ppp}$, which is considered a high intensity in PS. 

\begin{figure}[!tbh]
    \centering
    \includegraphics*[width=\columnwidth]{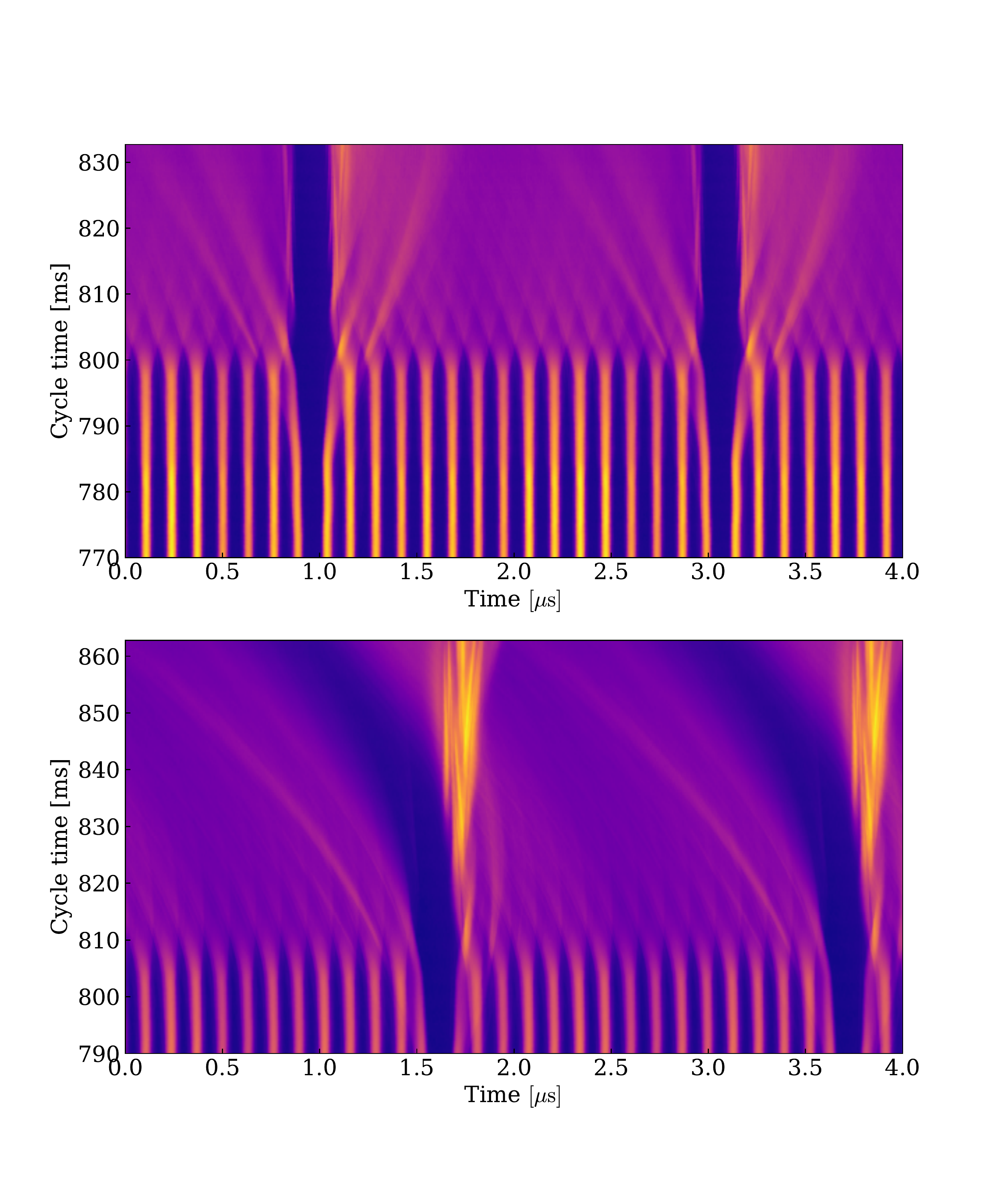}
    \caption{Longitudinal profile evolution with static~(top) and decelerating barrier~(bottom). The depth of the longitudinal gap clearly diminishes in the decelerating barrier bucket case.}
    \label{fig:longitudinal-profiles}
\end{figure}

\subsection{Accuracy of the prototype synchronisation with respect to extraction kick timing accuracy}

Initially the position of the kick to extract the core of the beam was measured with the barrier system disabled. A trace from the five-turn spill registered by a fast beam current transformer in the transfer line was low-pass filtered, and the peak position of the kick gap was detected in the longitudinal beam distribution. Despite the kicker timing being precise at the nanosecond level, the detected position can slightly vary from shot to shot depending on the longitudinal distribution in the islands. This causes a different kicker gap profile shape that will cause a different detected mean position. Hence the figure on the kick accuracy is also to be taken as the accuracy of the detection combined with the shot-to-shot variations of the beam conditions. As a practical consequence, it is reasonable to set the barrier width somewhat larger than the mean kicker rise time.

The longitudinal positions of the barriers were evaluated using the signal from the wall current monitor, similar to one acquisition in Fig.~\ref{fig:longitudinal-profiles}, in the PS just before extraction. Table~\ref{kick-barrier} shows the positions of the kick gap and barrier signals within the acquired oscilloscope traces from the moment of the trigger, which was repeateble at the nanosecond level. The standard deviation from the detected average position was calculated based on the number of traces corresponding to a cycle each, as indicated in the column on rightmost column.

\begin{table}[!hbt]
   \centering
   \caption{Detected kick gap and barrier position in the longitudinal profiles}
   \begin{tabular}{llll}
       \toprule
       Event & Position [$\SI{}{ns}$] & 
       $1\sigma$ std [$\SI{}{ns}$] & \# cycles \\
       \midrule
                   &     &  &  \\
            Core kick     & 10392 &  6.1  & 224 \\ 
            Barrier 1     &  402 & 10.8 & 247 \\ 
            Barrier 2      &  1453  &  10.7 & 247 \\
       \bottomrule
   \end{tabular}
   \label{kick-barrier}
\end{table}

In order to evaluate the reduction of the PS beam losses, similarly to the tests reported in~\cite{prab-barrier-bucket}, the septum that currently shields the extraction region during the rise time of the extraction kickers has to be retracted. Since the present tests were performed in parallel to the operational beams where the septum must be present for radiation protection reasons, these aspects can not be reported in this paper.

\section{CONCLUSIONS}

A barrier bucket PS-SPS synchronisation scheme was devised and commissioned with a technical demonstrator hardware connected to the present PS beam control. All elements of the synchronization have been validated with beam. The high-intensity beam in barrier buckets was extracted to SPS and delivered to North Area experiments. Further studies are needed to investigate intensity limits of the scheme. The present hard- and software implementation of the manipulation successfully guided the path to the full demonstration of the feasibility of the synchronised barrier-bucket transfer scheme. Several options, based on micro-controllers or programmable logic are under study for a future operational implementation. 

\section{ACKNOWLEDGEMENTS}
We would like to thank G. Papotti and S. Albright for the help with the SPS and PSB cycles and coordination. We would like to thank R. Piandani for the GTK~\cite{GTK} measurements at NA62~\cite{NA62}. We would also like to thank the SY-RF-FB section for the material and lab support. The studies would not have been possible without the dedicated, effective, and enthusiastic support of the PSB, PS and SPS operations teams. For the present version of the paper, one of the authors (MV) would like to thank D. Lens and F. Tamura for the feedback during the LLRF 2022 workshop.
%
%
	{%


\end{document}